# Measurement of atomic modulation direction using the azimuthal variation of first order Laue zone electron diffraction

Aurys Silinga[1], Christopher S. Allen[2,3], Juri Barthel[4], Colin Ophus[5], and Ian MacLaren[1,6],

1. SUPA School of Physics and Astronomy, University of Glasgow, Glasgow G12 8QQ, UK
2. electron Physical Science Imaging Centre, Diamond Light Source Ltd., OX11 0DE, UK
3. Department of Materials, University of Oxford, Parks Road, Oxford OX1 3PH, UK
4. Ernst Ruska-Centre (ER-C 2), Forschungszentrum Jülich GmbH, 52425 Jülich, Germany
5. NCEM, Molecular Foundry, Lawrence Berkeley National Laboratory, Berkeley 94720, USA
6. Ian MacLaren was an Editor of the journal during the review period of the article. To avoid a conflict of interest, he was blinded to the record and another editor processed this manuscript).

**Abstract**

We show that diffraction intensity into the First Order Laue Zone (FOLZ) of a crystal can have a strong azimuthal dependence, where this FOLZ ring appears solely because of unidirectional atom position modulation. Such a modulation was already known to cause the appearance of elliptical columns in atom resolution images, but we show that measurement of the angle via 4-dimensional Scanning Transmission Electron Microscopy (4DSTEM) is far more reliable and allows the measurement of the modulation direction with a precision of about 1° and an accuracy of about 3°. This method could be very powerful in characterising atomic structures in 3 dimensions by 4DSTEM, especially in cases where the structure is only found in nanoscale regions or crystals.

**Introduction**

Atomic resolution imaging using scanning transmission electron microscopy (STEM) has been crucially important to our understanding of materials and artificial heterostructures for the past 30 years. This started with the widespread appreciation that High Angle Annular Dark Field (HAADF) imaging was mainly incoherent and strongly dependent on the atomic number *Z* of the species being imaged (Pennycook and Jesson 1991, Hartel, Rose et al. 1996), and only became more critical after the introduction of aberration-correction (MacLaren and Ramasse 2014) . However, the majority of atomic-resolution imaging is purely concerned with a 2-dimensional projection of the sample, and not with the 3-dimensional structure, although inferences about modulation of atom positions along a column have been made from column shape analysis (Borisevich, Ovchinnikov et al. 2010, He, Ishikawa et al. 2015, Azough, Cernik et al. 2016). For this reason, analysis of column ellipticity has been included in open source analysis packages, such as *Atomap*, for quantification of atomic resolution images using Gaussian fitting (Nord, Vullum et al. 2017).

Higher-order Laue zone (HOLZ) rings occur at relatively high diffracted angles and, due to the curvature of the Ewald sphere, have components of the diffraction vector along the beam direction. HOLZ rings therefore reveal information about the 3D atomic structure by revealing lattice spacings along the beam direction (Jones, Rackham et al. 1977, Jesson and Steeds 1990, Spence and Koch 2001). This has been used in STEM imaging with a modified conventional annular detector to show a period tripling in sodium cobaltate (Huang, Gloter et al. 2010). More recently, the advent of fast, pixelated direct electron detectors and the 4DSTEM technique (Ophus 2019, MacLaren, Macgregor et al. 2020), has enabled high angle diffraction patterns to be acquired from every probe position on a scan. These patterns can then be analysed and quantified, leading to the development of high resolution HOLZ-STEM as a technique. This has been used to measure changes in crystal periodicity along the beam direction (Nord, Ross et al. 2019, Paterson, Webster et al. 2020), strain from the HOLZ ring radius (MacLaren, Macgregor et al. 2020), or atomic modulation at atomic resolution (Nord, Barthel et al. 2021). The present work exploits the azimuthal variation in the intensity of the First

Order Laue Zone (FOLZ) rings, which can be measured in certain crystals due to atomic position modulation.

**Methods**

A sample of double perovskite La$_2$CoMnO$_6$ (LCMO) was used in this work which had been grown on (111) LSAT (lanthanum aluminate - strontium aluminium tantalate (Chakoumakos, Schlom et al. 1998), a pseudocubic perovskite substrate with $a \approx 3.87$Å). Full details of growth and characterisation are given in (Kleibeuker, Choi et al. 2017). A specimen for transmission electron microscopy was prepared by a standard FIB liftout method, as also described in (Kleibeuker, Choi et al. 2017).

Scanning transmission electron microscopy was performed using a JEOL ARM300F at the ePSIC facility with a 40μm condensor aperture chosen to give a convergence angle of 29mrad and a nominal camera length of 8cm. Specifically, high angle annular dark field (HAADF) imaging was performed by recording a sequence of 8 images, to enable post-processing correction of sample drift and scan distortions (Sang and LeBeau 2014, Jones, Yang et al. 2015, Ophus, Ciston et al. 2016). In this case, each was sequentially rotated by 90° and the image sequences were aligned and summed using the scanning drift correction code of Ophus, Ciston et al. 2016 to produce high signal-to-noise distortion-free images. 4DSTEM datasets were recorded at a camera length short enough to allow diffracted angles up to ~100 mrad to be recorded on the detector (as in (MacLaren, Macgregor et al. 2020)) from the same areas. Nine such paired datasets were acquired with both good quality HAADF images and 4DSTEM datasets with diffraction patterns containing a HOLZ ring.

Refinement of atom positions was performed using Atomap (Nord, Vullum et al. 2017) and the ellipticity of each La-O atom column was determined.

Representative diffraction patterns were averaged from suitably chosen regions in the 4DSTEM datasets matching the areas used for ellipticity measurement and were transferred from a rectangular *x-y* representation to a polar $2\theta - \phi$ representation using *emilys* (https://github.com/ju-bar/emilys) after determination of a suitable pattern . An area in a certain radial range matching a HOLZ ring of interest was integrated to a 1D trace of HOLZ intensity versus $\phi$. This was then fitted to a suitable mathematical form using *scipy.optimize* within *python*.

*Jupyter* notebooks for all these operations will be archived together with the raw data for inspection and use by interested readers.

**Results and Discussion**

We performed STEM imaging along the <110>$_{cubic}$ direction of the double perovskite La$_2$CoMnO$_6$ (LCMO) on (111) LSAT. This sample is known to exhibit B-cation ordering, octahedral tilting, and expected to show atom modulation along La columns. LCMO has a P2$_1$/n structure which is almost orthorhombic with $a \approx b \approx \sqrt{2}a_{cubic}$ and $c \approx 2a_{cubic}$. Figure 1 is a schematic of the atomic structure along the [111] direction, as refined by (Bull, Playford et al. 2016) and shows the unidirectional position modulation of the La atoms, which would be a zigzag if seen from the side. This modulation is part of the symmetry reduction of the structure from primitive perovskite and doubles the periodicity of the lattice along this direction, which would be expected to give rise to an extra Laue zone (of radius $1/\sqrt{2}$ times that of the FOLZ in the primitive perovskite (Paterson, Webster et al. 2020)). The unidirectional modulation is of $\frac{1}{11.6}[010]$ in crystal coordinates or 0.47Å along the [010] direction. The component perpendicular to the [111] axis is about 0.2Å along the [$\bar{1}3\bar{1}$] direction, which makes an angle of 35° to the normal to the ($\bar{1}10$) plane (the vertical direction in Figure 1) (all standard crystallographic calculations from the structure of (Bull, Playford et al. 2016)).

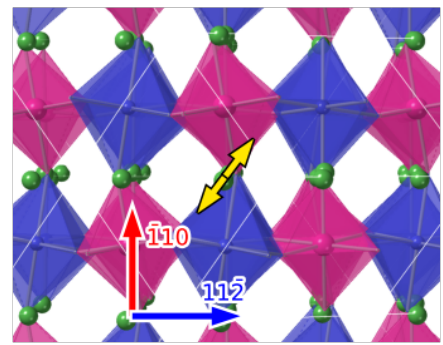

*Figure 1: Schematic diagram of the atomic structure of La$_2$CoMnO$_6$ with the [111] direction pointing out-of-plane towards the viewer. La atoms are green, Co are blue and Mn magenta, O are at corners of octahedra and not shown, the modulation direction of the La atoms is indicated with an orange arrow. The normals to the ($\bar{1}10$) and ($11\bar{2}$) planes are shown with red and blue arrows, respectively.*

An aligned HAADF-STEM sum image is shown in Figure 2a with arrows representing the evaluated ellipticity direction superimposed on La columns in the centre of the image. The average angle of ellipticity in this image was in the range of 20-38°, depending a little on the choice of fitting radius used for each atom column, which is discussed in more detail later.

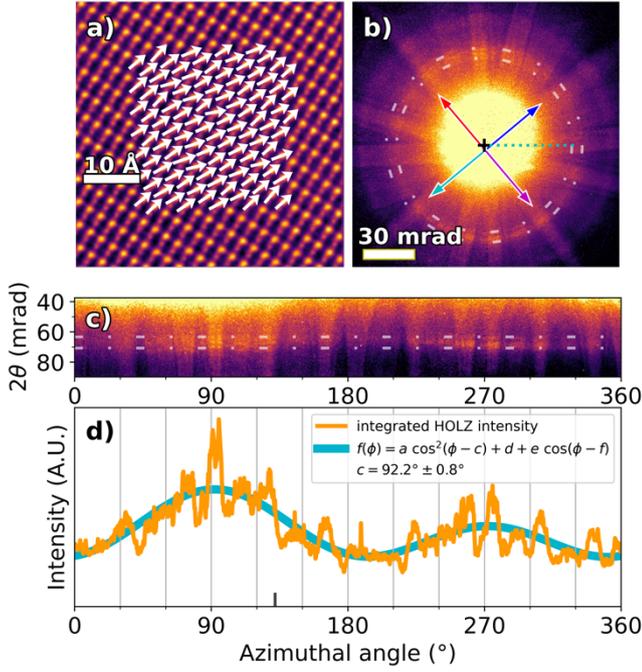

Figure 2. (a) HAADF atomic resolution image of modulated LCMO along a $\langle 111 \rangle_{cubic}$ direction with ellipticity arrows superimposed for some columns, (b) Average STEM diffraction pattern for this region, (c) Polar transformation of the FOLZ ring (d) Azimuthal variation of average FOLZ ring intensity, the peak intensity occurs at $c = 92.2 \pm 0.8°$. Figures 2b and 2c have had the contrast flattened by applying a power of 0.1 to the actual intensities to make visualisation of the real features at higher angles easier, but Figure 2d is calculated without any scaling of intensity.

For the same region, the diffraction pattern from a HOLZ-STEM dataset is shown in Figure 2b, with a clear FOLZ ring marked by the dotted lines. There is an obvious variation in intensity around the FOLZ ring, being highest at the top and bottom of the diffraction pattern whilst being almost invisible to the left and right. Figure 2c shows a polar transformation to a $2\theta - \phi$ description about the centre marked in 2b, with a calibration of $2\theta$ in mrad and $\phi$ in degrees. Again, dotted lines indicate the angular range in which most of the FOLZ intensity falls. These also help in pattern centre alignment, as any misjudgement thereof leads to the FOLZ not forming a straight line in the polar transform.

We integrated the area inside the dotted lines to create the plot in Figure 2d. This was then fitted to the empirical function (more on the justification thereof in the Supplemental Information):

$$I = a \cos^2(\phi - c) + d + e \cos(\phi - f) \quad [1]$$

The first term captures the azimuthal intensity variation on the ring arising from atom position modulation (with two-fold symmetry), where $a$ is the amplitude and $c$ the angular shift of the peak from 0° (defined in the range 0° ≤ $c$ < 180°). The second term and the $d$ parameter captures the offset of the ring from zero intensity due to background high angle scattering (mainly thermal diffuse scattering, as well as scattering from surface damage from the focused ion beam preparation (Nord, Barthel et al. 2021)). The third term (with one-fold symmetry) captures any small tilt away from the zone axis leading to higher intensity in one azimuthal direction. This mistilt gives a shift in the background intensity and means the centres of the zero order Laue zone (ZOLZ) and FOLZ are no longer concident, which does increase the intensity of the HOLZ ring on whichever side it occurs (towards 90° in this case). $e$ is the amplitude and $f$ is the angle of peak background intensity (defined in the range 0° ≤ $f$ < 360°). Only the $c$ term is significant to the present measurement as this gives the angle at which the FOLZ intensity peaks. This was measured as $c = 92.2 \pm 0.8°$ for the dataset shown in Figure 2. To compare this to the ellipticity measurement from HAADF-STEM, the relative rotation of the diffraction pattern to the STEM scan direction needs to be known. For the optical setup used in this experiment the relative rotation angle was measured to be 110° (with positive angles in the counter-clockwise direction), making this angle 202.2°, or 22.2° (because of the 180° period of the function), consistent with the measured ellipticity from the HAADF image (although referencing to crystal directions is more sensible in the end, as in the next paragraph). (Note, the 2-fold symmetry of the FOLZ is not present elsewhere in the diffraction pattern, for instance the second order Laue zone is almost isotropic in intensity).

The four arrows from the centre in Figure 2b show the normals to the $(\bar{1}10)$ (red-magenta) and $(11\bar{2})$ (blue-cyan) planes (using the Kikuchi lines for orientation, note plane normals are easier to measure directly in diffraction than crystal directions). There is a small deviation of the centre from the centre of the Laue zone due to slight sample mistilt from the $[111]$ zone axis. To get the rotation of the modulation from the $(\bar{1}10)$ plane normal, we subtracted the measured angle of the peak FOLZ intensity (92.2°) from the angle for the $(\bar{1}10)$ plane normal (the red arrow in the polar transform of Figure 2c) at 132°. This therefore measures the modulation direction as 37.2°±0.8° clockwise from the normal to $(\bar{1}10)$, in reasonable agreement with the structure refined by (Bull, Playford et al. 2016), where this angle was calculated as 35.1°.

Eight such measurements are shown in Figure 3 from different locations in the same sample (with some slight variations in thickness and crystal tilt across the specimen). We corrected the FOLZ data to crystal coordinates using the position of $(\bar{1}10)$ plane normal as seen in the polar transform. Error bars for each were calculated from the covariance of the $c$ parameter in equation 1. The diffraction-based measurement is consistent in all cases with an average of 38.1±1.0°. We corrected the HAADF-STEM data to the same crystal coordinate system using the position of the $(\bar{1}10)$ plane normal, which was determined as part of *Atomap* fitting (Nord, Vullum et al. 2017) to be at 56.9° (anticlockwise). All ellipticity measurements are smaller than this, meaning that all corrected measurements are clockwise rotations. These are plotted with an uncertainty estimated from the range of ellipse major axes given by *Atomap* fitting with different fit radii in the range of 25-45% of the nearest neighbour distance between atoms, combined with the random scatter in the measurement within each measured area (which was typically large). The average of all these measurements was a rotation of 40.2±10.2° clockwise. Clearly, this measurement is broadly consistent with the diffraction-based measurement and the crystallographic theory but there is much higher scatter in the measurement when using the ellipticity from HAADF images. This could be due to a number of factors. Details of the alignment procedure were tested by fitting with different codes, using different areas in the same image and this was found to be insignificant. The radius around the intensity maximum chosen for the ellipse fitting (as fractions of atom spacing) was found to be significant, and the apparent angle increased with increasing radius, probably because of the influence of tails of intensity from neighbouring B-site atom columns. It is possible also that residual microscope aberrations and sample mistilt have an effect and it is well known that even with zero drift of astigmatism and coma with time, the apparent aberrations can vary from one sample region to another. However, slow drift of coma and astigmatism with time are also often noted, and could be having an effect. We therefore conclude that measurements of ellipticity from an image based method are less accurate than the diffraction based methods (large scatter and possible systematic errors on image based measures of ellipticity was a problem previously noted when analysing HAADF-STEM data on a modulated atom column in a different material (Azough, Cernik et al. 2016) but not commented on in print at the time).

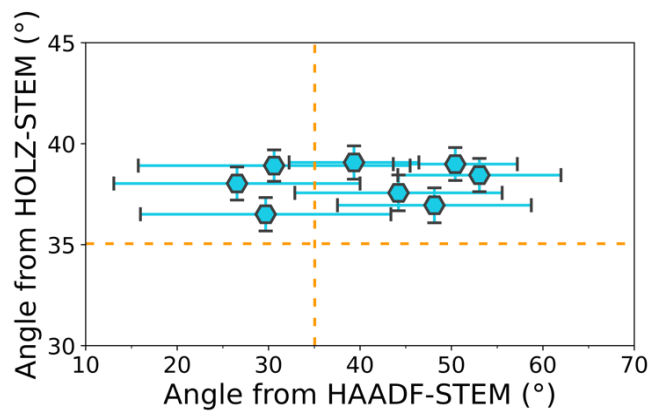

*Figure 3. Correlation of atom modulation measurements from HOLZ-STEM and HAADF-. All angles are clockwise rotations from the $(\bar{1}10)$ plane normal. The expectations from the refinement of Bull et al.[19] are shown as orange dotted lines. Error bars are calculated from the standard deviation of each measurement, and including an uncertainty estimate from the variation as a result of fitting radius for the HAADF-STEM measurement.*

The azimuthal variation on the FOLZ ring seen in the diffraction patterns (as exemplified by Figure 2b) agrees well with expectations from simulation. Figure 4a shows a simplistic kinematic calculation of diffracted intensities (which was calculated using *py4DSTEM* (Savitzky, Zeltmann et al. 2021)) for a low convergence angle probe passing through the crystal in the same orientation as represented in Figure 1. There is a peak in FOLZ intensity along the in-plane component of the modulation direction, [010], where the spacings of the planes responsible for reflections in the FOLZ would be most affected by La modulation along the beam direction (e.g. the $\bar{5}, 13, \bar{7}$ reflection in Figure 4a). In contrast to this, there is very little intensity perpendicular to this (e.g. the very weak $12, 1, \overline{12}$ spot indicated) where the plane spacing is almost unaffected by the La-modulation as the modulation is primarily within the diffracting planes, and structure factor will therefore remain very small. This shows that simple kinematic calculations are enough to explain the existence of the effect. Nevertheless, a more realistic simulation matching the experimental conditions was calculated by a multislice method using *Dr. Probe* (Barthel 2018) (full details in *Supplemental Materials*) and is shown in Figure 4b, showing the same azimuthal modulation, with the same peak and trough directions for intensity (as discussed by (Nord, Barthel et al. 2020), there are reasons why simulations appear sharper and with lower background than experiment). Thus, whilst simple kinematic arguments using structure factors are quite illustrative (and a more detailed argument is presented in the Supplemental Information), the same trend holds in a fully dynamical calculation. This also implies that the effect would be visible over a wide thickness range, from thin where plural scattering is almost negligible, to thicker where full dynamical treatment is essential. Applying the same measurement as in Figure 2 on the simulated data in Figure 4b results in a calculated angle of 37.7±1.2°, in excellent agreement with the experimental measurement above of 38.1±0.8° (see *Supplemental Materials* for full details of this measurement).

It seems that there is a small discrepancy between a measurement of a 38.1° rotation of modulation direction from the $(\bar{1}10)$ plane normal and a crystallographic calculation of 35.1°. The fact that a similar discrepancy (of 37.7°) persists in simulations suggests that some feature of the measurement introduces a systematic error, for reasons that are not entirely clear. Possibly, the appearance of specific Kikuchi bands or reflections at specific angles introduces a slight bias away from the true direction (as the brightest parts on the FOLZ ring are at intersections with Kikuchi bands). As such, this method lacks the precision of refinement from neutron diffraction from bulk crystal (Bull, Playford et al. 2016). Nevertheless, this has the major advantage of having a spatial resolution of nanometres or better. Moreover, a model-based approach of comparison of experiment with simulation can improve the quantitative interpretation. This would certainly be able to identify cases where the structure in a thin film deviates from the bulk structure, as previously noted for $LaFeO_3$ (Nord, Ross et al. 2018), and thereby provide enough information to constrain a new DFT refinement. In the present case, it can be confirmed with the aid of the simulation that the thin film structure shows no detectable deviation from the bulk structure for $La_2CoMnO_6$ previously refined by (Bull, Playford et al. 2016).

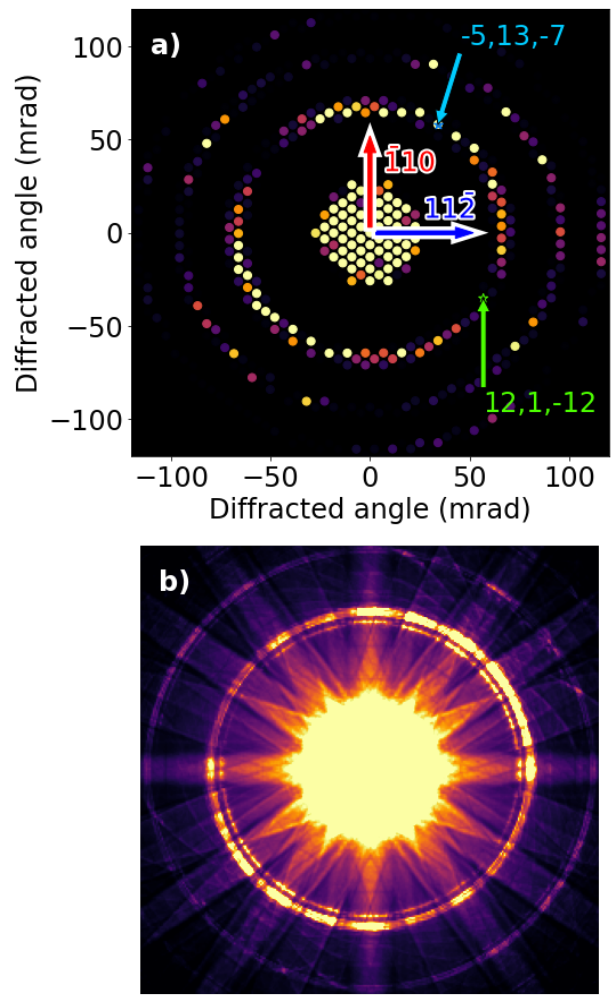

*Figure 4: Simulated diffraction patterns of the structure shown in Figure 1 (with same crystal orientation): a) kinematic calculation of spot intensities with low convergence angle to allow individual diffraction spots to be distinguished, b) full multislice calculation for beam convergence angle close to that used in the experimental study.*

Whilst the simulations were done with the best approaches known to the authors (Barthel, Cattaneo et al. 2020, Nord, Barthel et al. 2021), the geometry and intensity of the FOLZ is not strongly dependent on such detail. Changing convergence angle would not change the occurrence of the effect (c.f. Figure 4), only the real space resolution of 4DSTEM through the Abbe criterion. Changing kV would affect diffracted angles and relative intensities, but would not affect the actual azimuthal intensity variation either. The datasets examined here covered a range of sample thicknesses from below 50 nm to above 100 nm, and similar results were seen in all cases. This likely only fails where the crystal is too thin and the structure is lost due to damage, or too thick and separating elastic FOLZ signal from diffuse background scattering becomes harder due to the signal-to-background ratio becoming poor. Thus, this method would be robust and applicable in any STEM with a suitable detector and acquisition hardware for 4DSTEM, and kV and probe angle could be adjusted to user requirements for their system and sample.

It should be noted that the azimuthal variation seen here would be expected also in selected area diffraction (SAD) and convergent beam electron diffraction patterns (CBED) in conventional TEM. The first would only work provided there were a crystal showing the effect over hundreds of nanometres (rather than the 50 nm film thickness here). CBED would certainly work in this case, but the correlation to HAADF imaging in STEM would be harder. The major advantage of 4DSTEM is that not only is correlation of imaging and diffraction straightforward, but the spatially resolved mapping of FOLZ directionality would also be possible, and likely at atomic resolution (c.f. (Nord, Barthel et al. 2021)). Spatially resolved mapping of this azimuthal dependence of FOLZ intensity will be investigated further in a future manuscript.

## Conclusions

In conclusion, we show that the first order Laue zone of an ordered structure can have a strong, approximately sinusoidal azimuthal intensity dependence, and that this correlates with and therefore reveals the direction of the atom modulation that gave rise to it. Measurement of modulation direction via the azimuthal intensity variation in diffraction is both easier and more reliable than column shape analysis from atomic resolution imaging. Such azimuthal variation is also visible in other materials, such as Figure 3 in our previous work (Nord, Barthel et al. 2021). Whilst this azimuthal variation still lacks some of the three-dimensional information available in a good neutron diffraction refinement (Bull, Playford et al. 2016), this technique opens up new possibilities for three dimensional studies of crystallography on the nanoscale, which therefore serves as a valuable complement to neutron or X-ray diffraction. These new possibilities could include spatial variations in atomic modulation direction and strength in response to strain or chemistry, such as within epitaxial heterostructures and core-shell structures.

## Acknowledgements

We thank Diamond Light Source for access and support in use of the electron Physical Science Imaging Centre (Instrument E02, proposal number MG27893) that contributed to the results presented here . Work at the Molecular Foundry was supported by the Office of Science, Office of Basic Energy Sciences, of the U.S. Department of Energy under Contract No. DE-AC02-05CH11231. The authors gratefully acknowledge Steven Zeltmann at the Lawrence Berkeley National Laboratory for assistance with *py4DSTEM*. The sample was provided by Dr J.E. Kleibeuker and Prof. J.L. Driscoll of the University of Cambridge. The first version of the code for calculating Figure 4a was initially developed in an undergraduate research project with the aid of Mr Joss Thompson.

## Supplemental Information and Data

A supplemental information file is presented along with this paper giving more details of data processing and simulation methods.

*Jupyter* notebooks and all the raw data needed for them to generate the figures in this paper are archived at http://dx.doi.org/10.5525/gla.researchdata.1470 so the methods used in this work are accessible (even if the functions are not fully polished code suitable for release as a library at this point).

## Conflict of Interest

The authors declare none.

## References

Azough, F., R. I. Cernik, B. Schaffer, D. Kepaptsoglou, Q. M. Ramasse, M. Bigatti, A. Ali, I. MacLaren, J. Barthel, M. Molinari, J. D. Baran, S. C. Parker and R. Freer (2016). Tungsten Bronze Barium Neodymium Titanate ($Ba_{6-3n}Nd_{8+2n}Ti_{18}O_{54}$): An Intrinsic Nanostructured Material and Its Defect Distribution. Inorganic Chemistry 55, 3338-3350.

Barthel, J. (2018). Dr. Probe: A software for high-resolution STEM image simulation. Ultramicroscopy 193, 1-11.

Barthel, J., M. Cattaneo, B. G. Mendis, S. D. Findlay and L. J. Allen (2020). Angular dependence of fast-electron scattering from materials. Physical Review B 101, 184109.

Borisevich, A., O. S. Ovchinnikov, H. J. Chang, M. P. Oxley, P. Yu, J. Seidel, E. A. Eliseev, A. N. Morozovska, R. Ramesh, S. J. Pennycook and S. V. Kalinin (2010). Mapping Octahedral Tilts and Polarization Across a Domain Wall in BiFeO3 from Z-Contrast Scanning Transmission Electron Microscopy Image Atomic Column Shape Analysis. ACS Nano 4, 6071-6079.

Bull, C. L., H. Y. Playford, K. S. Knight, G. B. G. Stenning and M. G. Tucker (2016). Magnetic and structural phase diagram of the solid solution $LaCo_xMn_{1-x}O_3$. Physical Review B 94, 014102.

Chakoumakos, B. C., D. G. Schlom, M. Urbanik and J. Luine (1998). Thermal expansion of LaAlO3 and (La,Sr)(Al,Ta)O3, substrate materials for superconducting thin-film device applications. Journal of Applied Physics 83, 1979-1982.

Hartel, P., H. Rose and C. Dinges (1996). Conditions and reasons for incoherent imaging in STEM. Ultramicroscopy 63, 93-114.

He, Q., R. Ishikawa, A. R. Lupini, L. Qiao, E. J. Moon, O. Ovchinnikov, S. J. May, M. D. Biegalski and A. Y. Borisevich (2015). Towards 3D Mapping of $BO_6$ Octahedron Rotations at Perovskite Heterointerfaces, Unit Cell by Unit Cell. Acs Nano 9, 8412-8419.

Huang, F. T., A. Gloter, M. W. Chu, F. C. Chou, G. J. Shu, L. K. Liu, C. H. Chen and C. Colliex (2010). Scanning Transmission Electron Microscopy Using Selective High-Order Laue Zones: Three-Dimensional Atomic Ordering in Sodium Cobaltate. Physical Review Letters 105, 125502.

Jesson, D. E. and J. W. Steeds (1990). An investigation of three-dimensional diffraction from 2Hb-MoS2. Philosophical Magazine A 61, 363-384.

Jones, L., H. Yang, T. J. Pennycook, M. S. J. Marshall, S. Van Aert, N. D. Browning, M. R. Castell and P. D. Nellist (2015). Smart Align-a new tool for robust non-rigid registration of scanning microscope data. Advanced Structural and Chemical Imaging 1.

Jones, P. M., G. M. Rackham and J. W. Steeds (1977). Higher-Order Laue Zone Effects in Electron-Diffraction and their use in Lattice-Parameter Determination. Proceedings of the Royal Society of London Series a-Mathematical Physical and Engineering Sciences 354, 197-222.

Kleibeuker, J. E., E.-M. Choi, E. D. Jones, T.-M. Yu, B. Sala, B. A. MacLaren, D. Kepaptsoglou, D. Hernandez-Maldonado, Q. M. Ramasse, L. Jones, J. Barthel, I. MacLaren and J. L. MacManus-Driscoll (2017). Route to achieving perfect B-site ordering in double perovskite thin films. NPG Asia Materials 9.

MacLaren, I., T. A. Macgregor, C. S. Allen and A. I. Kirkland (2020). Detectors—The ongoing revolution in scanning transmission electron microscopy and why this important to material characterization. APL Materials 8, 110901.

MacLaren, I. and Q. M. Ramasse (2014). Aberration-corrected scanning transmission electron microscopy for atomic-resolution studies of functional oxides. International Materials Reviews 59, 115-131.

Nord, M., J. Barthel, C. S. Allen, D. McGrouther, A. I. Kirkland and I. MacLaren (2020). Atomic resolution HOLZ-STEM imaging of atom position modulation in oxide heterostructures. submitted to Ultramicroscopy.

Nord, M., J. Barthel, C. S. Allen, D. McGrouther, A. I. Kirkland and I. MacLaren (2021). Atomic resolution HOLZ-STEM imaging of atom position modulation in oxide heterostructures. Ultramicroscopy 226, 113296.

Nord, M., A. Ross, D. McGrouther, J. Barthel, M. Moreau, I. Hallsteinsen, T. Tybell and I. MacLaren (2018). 3D sub-nanoscale imaging of unit cell doubling due to octahedral tilting and cation modulation in strained perovskite thin films. ArXiV, 1810.07501.

Nord, M., A. Ross, D. McGrouther, J. Barthel, M. Moreau, I. Hallsteinsen, T. Tybell and I. MacLaren (2019). Three-dimensional subnanoscale imaging of unit cell doubling due to octahedral tilting and cation modulation in strained perovskite thin films. Physical Review Materials 3, 063605.

Nord, M., P. E. Vullum, I. MacLaren, T. Tybell and R. Holmestad (2017). Atomap: a new software tool for the automated analysis of atomic resolution images using two-dimensional Gaussian fitting. Advanced Structural and Chemical Imaging 3, 9.

Ophus, C. (2019). Four-Dimensional Scanning Transmission Electron Microscopy (4D-STEM): From Scanning Nanodiffraction to Ptychography and Beyond. Microscopy and Microanalysis 25, 563-582.

Ophus, C., J. Ciston and C. T. Nelson (2016). Correcting nonlinear drift distortion of scanning probe and scanning transmission electron microscopies from image pairs with orthogonal scan directions. Ultramicroscopy 162, 1-9.

Paterson, G. W., R. W. H. Webster, A. Ross, K. A. Paton, T. A. Macgregor, D. McGrouther, I. MacLaren and M. Nord (2020). Fast Pixelated Detectors in Scanning Transmission Electron Microscopy. Part II: Post-Acquisition Data Processing, Visualization, and Structural Characterization. Microscopy and Microanalysis 26, 944-963.

Pennycook, S. J. and D. E. Jesson (1991). High-Resolution Z-Contrast Imaging of Crystals. Ultramicroscopy 37, 14-38.

Sang, X. and J. M. LeBeau (2014). Revolving scanning transmission electron microscopy: Correcting sample drift distortion without prior knowledge. Ultramicroscopy 138, 28-35.

Savitzky, B. H., S. E. Zeltmann, L. A. Hughes, H. G. Brown, S. Zhao, P. M. Pelz, T. C. Pekin, E. S. Barnard, J. Donohue, L. Rangel DaCosta, E. Kennedy, Y. Xie, M. T. Janish, M. M. Schneider, P. Herring, C. Gopal, A. Anapolsky, R. Dhall, K. C. Bustillo, P. Ercius, M. C. Scott, J. Ciston, A. M. Minor and C. Ophus (2021). py4DSTEM: A Software Package for Four-Dimensional Scanning Transmission Electron Microscopy Data Analysis. Microscopy and Microanalysis 27, 712-743.

Spence, J. C. H. and C. Koch (2001). On the measurement of dislocation core periods by nanodiffraction. Philosophical Magazine B-Physics of Condensed Matter Statistical Mechanics Electronic Optical and Magnetic Properties 81, 1701-1711.

# Supplemental Information - Measurement of directional atomic modulation direction using the azimuthal variation of first order Laue zone electron diffraction

Aurys Silinga[1], Christopher S. Allen[2,3], Juri Barthel[4], Colin Ophus[5], Ian MacLaren[1],
1. SUPA School of Physics and Astronomy, University of Glasgow, Glasgow G12 8QQ, UK
2. electron Physical Science Imaging Centre, Diamond Light Source Ltd., OX11 0DE, UK
3. Department of Materials, University of Oxford, Parks Road, Oxford OX1 3PH, UK
4. Ernst Ruska-Centre (ER-C 2), Forschungszentrum Jülich GmbH, 52425 Jülich, Germany
5. NCEM, Molecular Foundry, Lawrence Berkeley National Laboratory, Berkeley 94720, USA

## Microscopy and data-processing methods

The LCMO sample was imaged using the E02 instrument in the EPSIC facility at Diamond light source. This is a JEOL ARM300F, and was operated at 200 kV, with a probe convergence angle of ~29 mrad, both to record HAADF images and to record 4DSTEM datasets to the Merlin EM detector (a 2x2 tiled detector mounted below the viewing screen).

HAADF images were recorded as 8 sequential images and then aligned and summed using the software of Ophus *et al*.[1]. As stated in the manuscript, atom positions were found and fitted, and the ellipticity determined using https://atomap.org/. Normal mean and standard deviation for the ellipticity direction was determined using standard *numpy* functions.

4DSTEM datasets containing large angular range diffraction patterns were processed by the following procedure:

- Datasets were loaded with hyperspy (https://hyperspy.org/) and filtered to remove dead and hot pixels (https://pixstem.org/using_pixelated_stem_class.html#finding-and-removing-bad-pixels), and then resaved

- Nine image areas were defined in each dataset for analysis in a 3x3 grid and the sum diffraction pattern calculated for each of the nine.

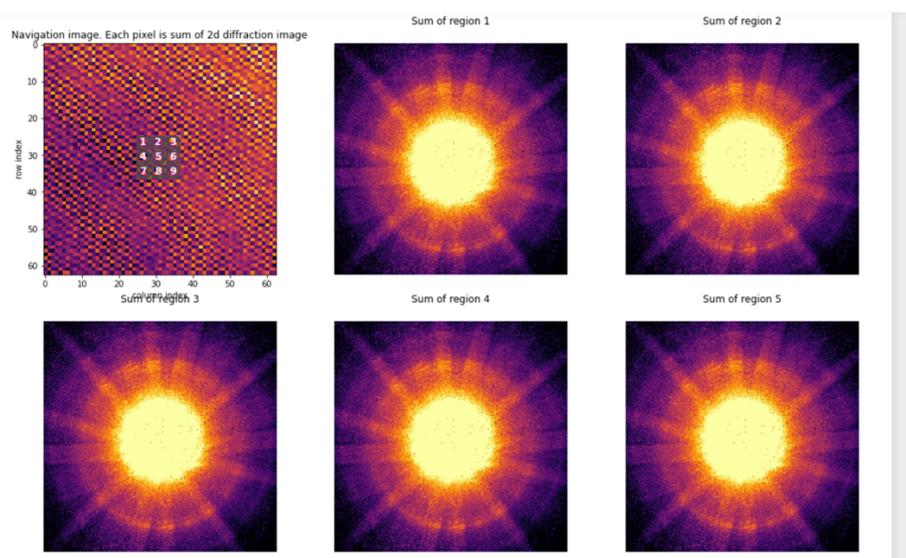

Figure S1: Calculation of average diffraction patterns from small areas in the 4DSTEM dataset (cropped version – only first five of the nine shown).

- The polar transform about the pattern centre (found by trial and error) was calculated using the *Emilys* package (https://github.com/ju-bar/emilys). At the radius of the ring, (about 180 pixels), it was chosen to sample the azimuthal direction in $\approx 2\pi r$ samples to keep a similar azimuthal sampling to the pixel size, which meant us choosing to use a sampling of 1080 pixels (i.e. 1/3° divisions).

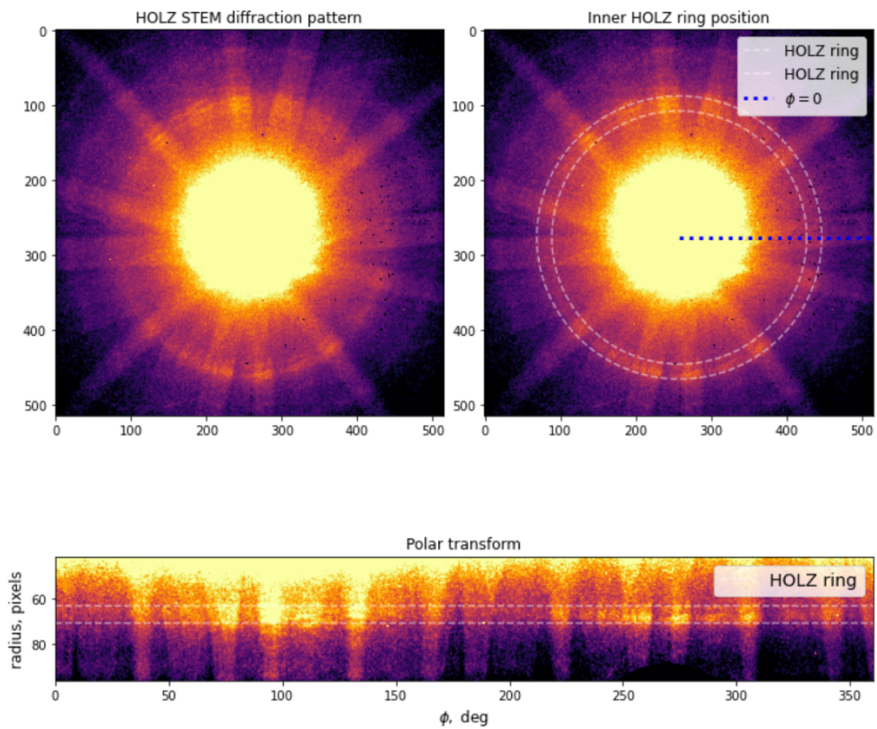

Figure S2: Polar transform of one diffraction pattern, after finding the centre of the FOLZ.

- The resulting polar transformed diffraction patterns were summed into 1D line profiles in a width of ±10 pixels about the ideal radius

- This was fitted using *scipy.optimize* to the function described, and the angle shift parameter for the $\cos^2$ function was extracted.

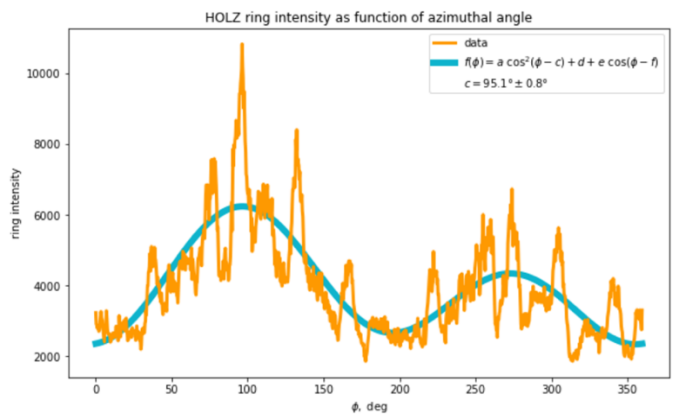

a,c,d,e,f = [2701.2518541077807, 95.05369498155156, 2585.2987953444012, 963.3197574850659, 104.89092912797996] +/-
[77.56385591226567, 0.8225966775575233, 47.49796735935611, 38.781927883380526, 2.306649249974957]
region 2

Fig S3: 1D sum of the FOLZ, fit plot, and fit parameters.

- Diffraction to image rotation was determined by scanning a defocused probe with a recognisable feature in a shadow image, and then rotating the diffraction pattern in the dataset until moving along the two orthogonal scan axes moves the shadow image in the correct directions.

**Full results**

A full table of results is presented below in Table S1 for the different datasets used, which contained a pair of a good HAADF image and a good 4DSTEM dataset.

| | Fitting radius (as a fraction of nearest neighbour distance) | | | | | |
|---|---|---|---|---|---|---|
| Timestamp | 0.25 | 0.30 | 0.35 | 0.40 | 0.45 | FOLZ |
| 154716 | 48.7 ± 11.0 (1.22) | 47.7 ± 9.9 (1.21) | 45.5 ± 9.9 (1.19) | 42.1 ± 10.0 (1.17) | 37.0 ± 10.7 (1.16) | 37.6 ± 0.9 |
| 160139 | 34.7 ± 15.5 (1.16) | 32.7 ± 13.7 (1.15) | 30.4 ± 12.8 (1.15) | 27.2 ± 11.8 (1.15) | 23.3 ± 10.8 (1.15) | 36.5 ± 0.8 |
| 160712 | 39.5 ± 16.6 (1.19) | 35.3 ± 14.2 (1.18) | 31.2 ± 12.9 (1.18) | 25.9 ± 11.2 (1.19) | 21.0 ± 9.8 (1.2) | 38.9 ± 0.8 |
| 161216 | 31.2 ± 17.3 (1.18) | 28.8 ± 14.0 (1.19) | 26.9 ± 12.1 (1.2) | 24.3 ± 10.9 (1.21) | 21.5 ± 10.2 (1.23) | 38.0 ± 0.8 |
| 162329 | 41.0 ± 8.1 (1.31) | 40.8 ± 6.8 (1.31) | 40.2 ± 6.3 (1.31) | 38.6 ± 6.2 (1.29) | 36.1 ± 6.5 (1.28) | 39.1 ± 0.8 |
| 162847 | 50.5 ± 7.2 (1.33) | 50.8 ± 6.5 (1.33) | 50.8 ± 6.4 (1.31) | 50.4 ± 6.6 (1.29) | 49.6 ± 7.1 (1.27) | 39.0 ± 0.8 |
| 163337 | 56.2 ± 10.7 (1.25) | 55.0 ± 8.2 (1.24) | 53.4 ± 7.7 (1.23) | 51.6 ± 7.5 (1.22) | 49.0 ± 8.2 (1.2) | 38.4 ± 0.8 |
| 163801 | 51.8 ± 9.4 (1.26) | 50.6 ± 9.5 (1.24) | 48.9 ± 9.6 (1.21) | 46.3 ± 10.3 (1.19) | 42.9 ± 11.1 (1.17) | 37.0 ± 0.9 |

Table S1: Ellipticity data from Atomap with FOLZ azimuthal peak position data. Ellipticity angles with standard deviations for each dataset (by timestamp) are given for five different fitting radii. The actual ellipticity (ratio of major and minor axes) is in brackets in each case. The relatively small ellipticity values may be contributing to the high random uncertainties.

**Reason for the azimuthal modulation of FOLZ intensity– a kinematical argument**

As a super simple model to illustrate the essential origins of the extra Laue zone, we just consider the scattering from two atoms, which could easily be extended to a single string of atoms (not in a 3D crystal). We consider elastic electron scattering from a target with period length $c$ consisting of two identical scattering objects (atoms) per period for an incident plane wave with wavevector $\mathbf{k} = k(0,0,1)$ along the $z$ direction, $k = 1/\lambda$, and $\lambda$ the De Broglie wavelength of the electron. The outgoing wave vector for elastic scattering is denoted by $\mathbf{k}' = k(\sin\theta \cos\phi, \sin\theta \sin\phi, \cos\theta)$ with a scattering angle $\theta$, as illustrated in Fig. 1a. The azimuthal angle $\phi$ is omitted in the drawing.

We first consider a setup in which the two atoms are aligned perfectly with the incident wave vector and the distance between the two atoms is half the period length $c$ with a distance vector of $\mathbf{R} = c/2\,(0,0,1)$. A path difference of $\Delta s = \lambda \mathbf{R}(\mathbf{k} - \mathbf{k}')$ for the two scattered waves results in general in an intensity including interference according to:

$$I = 2|A|^2 \left(1 + \cos\left[\tfrac{2\pi}{\lambda}\Delta s\right]\right) = 4|A|^2 \cos^2\left[\tfrac{\pi}{\lambda}\Delta s\right] \quad (1)$$

where $A$ is the amplitude of scattering from each of the identical atoms.

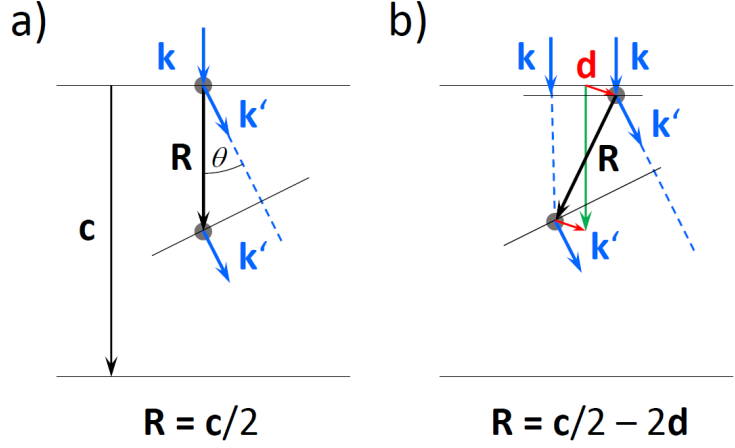

Fig S4: Schematic diagram of scattering of a plane wave by two atoms (a) aligned in a column with period length $c$, of twice the atom distance along the incident wave vector **k** and (b) including an antisymmetric displacement by a vector **d**. The difference in distance vectors **R** between the atoms leads to different optical path lengths for waves scattered at the two atoms to an outgoing wave vector **k'** with angle $\theta$ to the z axis.

In the scenario depicted in Fig. 1a, where the two atoms are periodically aligned along the incident wave vector, extinction of the intensity in eq. (1) is obtained for $\Delta s = \lambda\left(m + \tfrac{1}{2}\right)$, where $m$ is an integer. Considering the first of these multiple cases of extinction ($m = 0$), this leads to a scattering condition of:

$$1 = 2\mathbf{R}(\mathbf{k} - \mathbf{k'}) = c\frac{(1-\cos\theta)}{\lambda} \quad (2)$$

This condition determines the scattering angle $\theta$ at which we expect the first-order Laue zone to appear when the two atoms are displaced, as illustrated in Fig. 1b.

We consider now an anti-symmetric displacement of the two atoms by vectors $\mathbf{d} = d(\sin\alpha\cos\beta, \sin\alpha\sin\beta, \cos\alpha)$ and $-\mathbf{d}$ to the respective periodic positions in the first case. The displacement vector **d** is assumed to be small compared to $c$ but can be oriented arbitrarily with an angle $\alpha$ to the z-axis and angle $\beta$ to the x-axis. In this case, the distance vector is $\mathbf{R} = c/2\,(0,0,1) - 2\mathbf{d}$ and the path difference is given by:

$$\Delta s = \tfrac{c}{2}(1 - \cos\theta) - 2d(1-\cos\theta)\cos\alpha + 2d\sin\theta\sin\alpha\cos[\phi - \beta] \quad (3)$$

By applying the scattering condition of eq. (2) in eq. (3), we obtain:

$$\Delta s = \lambda\left(\tfrac{1}{2} - \tfrac{2d}{c}\cos\alpha + \tfrac{2d}{c}\sqrt{\tfrac{2c}{\lambda} - 1}\sin\alpha\cos[\phi - \beta]\right) \quad (4)$$

and inserting eq. (4) in eq. (1) gives the intensity:

$$I = 4|A|^2 \sin^2\left[\tfrac{2\pi d}{c}\left(\cos\alpha + \sqrt{\tfrac{2c}{\lambda} - 1}\sin\alpha\cos[\phi - \beta]\right)\right] \quad (5)$$

For a finite displacement $d > 0$, eq. (5) predicts a non-zero intensity at the scattering angle $\cos\theta$., i.e. the generation of a first-order Laue zone. For displacements with a component in the x/y-plane, i.e. for $0 < \alpha \leq \pi$, we expect an intensity variation in the azimuthal direction of the diffraction pattern, relative to the projected displacement vector by the term containing $\cos[\phi - \beta]$. This is the phenomenon measured and discussed in the paper.

The first-order term of a Taylor series of the sin function in eq. (5) for small fractions $\frac{d}{c}$ gives:

$$I = 16|A|^2 \left(\frac{d}{c}\right)^2 \left(\cos^2\alpha - 2\sqrt{\frac{2c}{\lambda} - 1}\cos\alpha\cos[\phi - \beta] + \left[\frac{2c}{\lambda} - 1\right]\sin^2\alpha\cos^2[\phi - \beta]\right) (6)$$

in which we can identify the three terms of the empirical fitting function applied in the analysis in the main paper. Specifically, the first term is a constant just depending on $\alpha$ (although other constant terms also apply to the full diffraction pattern, principally from thermal diffuse scattering). The second term is a 1-fold function of $\phi - \beta$. The third term is a two-fold function of $\phi - \beta$ (i.e. peaking twice in a $2\pi$ azimuthal rotation). Note, that the twofold part function is maximised for a **d** vector perpendicular to the beam direction with $\alpha = \frac{\pi}{2}$, but would be zero for $\alpha = 0$ (ie displacements purely parallel to beam direction). This emphasises that the effect is an effective measure of transverse modulations along atom columns to the beam direction. This illustrates that the effect does not require any complex considerations of dynamical scattering to understand its origins, even if correctly accounting for the details of intensity distribution for any specific set of experimental conditions would certainly require high quality simulation based on robust theory involving dynamical scattering.

## Simulation of diffraction patterns

The simple disc diffraction pattern of Figure 4a was calculated using intensities from a simple kinematic calculation in *py4dstem*[2] plotted using custom plotting code using *matplotlib*.

The dynamical calculation of Figure 4b was calculated using Dr Probe[3] using the following parameters:

- Acceleration voltage        200 kV
- Beam direction              [111]
- Convergence semiangle       20 mrad
- Sample thickness            50 nm
- Angular range simulated     120 mrad
- Sampling in diffraction plane   384×384 pixels
- Pixel size                  0.8068 mrad/pixel
- Calculation precision       float-32
- Structure                   Bull *et al.*[4]

Thermal diffuse scattering was included in the simulation according to the quantum excitation of phonons model[5] and the Einstein model of uncorrelated thermal vibration. Accordingly, several hundred random positional configurations of the atomic structure were generated to reproduce the thermal mean squared displacement parameters given in the structure model, and several hundred multislice passes were averaged each with a different random permutation of positional configurations.

## Polar transform and fitting of the simulated pattern

Polar transform was performed on the simulated data using the exact same code as for the experimental data yielding the results shown in Figure S4.

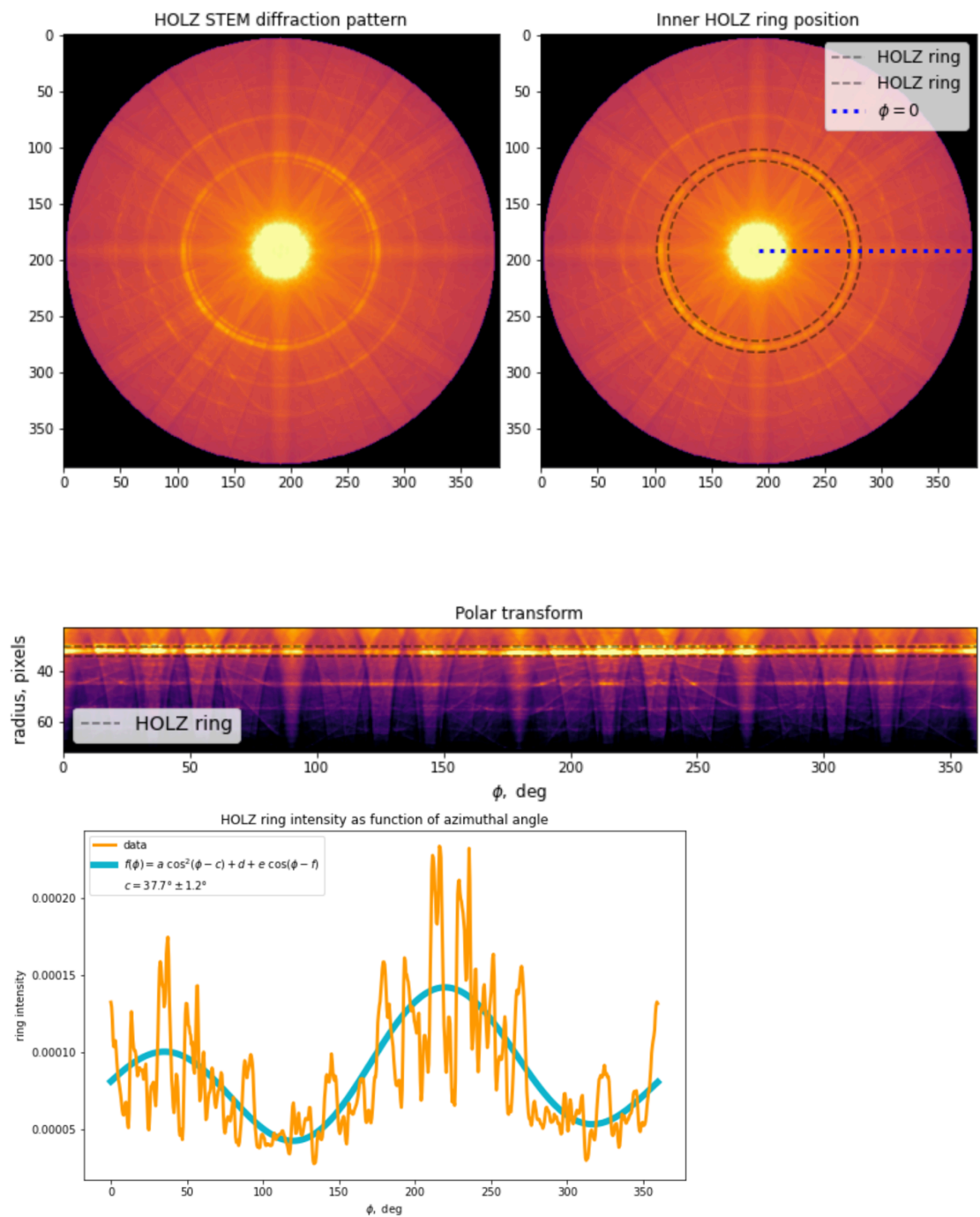

a,c,d,e,f = [7.157274082514289e−05, 37.74813102524245, 4.934995974633684e−05, 2.1565477985098358e−05, 232.39673499923148] +/− [2.923141209505613e−06, 1.1700240366623762, 1.790051102347615e−06, 1.461570604752765e−06, 3.8831426391356123]

Fig S4: Polar transform of the simulated diffraction pattern and the fitting of the azimuthal intensity modulation therein to give a peak intensity at 37.7°.

# References

1. C. Ophus, J. Ciston and C. T. Nelson, Ultramicroscopy **162**, 1-9 (2016).
2. B. H. Savitzky, S. E. Zeltmann, L. A. Hughes, H. G. Brown, S. Zhao, P. M. Pelz, T. C. Pekin, E. S. Barnard, J. Donohue, L. Rangel DaCosta, E. Kennedy, Y. Xie, M. T. Janish, M. M. Schneider, P. Herring, C. Gopal, A. Anapolsky, R. Dhall, K. C. Bustillo, P. Ercius, M. C. Scott, J. Ciston, A. M. Minor and C. Ophus, Microsc. Microanal. **27** (4), 712-743 (2021).
3. J. Barthel, Ultramicroscopy **193**, 1-11 (2018).
4. C. L. Bull, H. Y. Playford, K. S. Knight, G. B. G. Stenning and M. G. Tucker, Phys. Rev. B **94** (1), 014102 (2016).
5. B. D. Forbes, A. V. Martin, S. D. Findlay, A. J. D'Alfonso and L. J. Allen, Phys. Rev. B **82** (10), 104103 (2010).